\documentclass{elsart}
\usepackage{graphicx}

\begin{document}
\begin{frontmatter}

\title{Short-Range Structural Transformations\\ in Water at High Pressures}

\author[label1]{Ramil M. Khusnutdinoff}
\ead{khrm@mail.ru}
\author[label1]{Anatolii V. Mokshin}
\address[label1]{Department of Physics, Kazan (Volga region) Federal University,
Kremlevskaya Street 18, 420008 Kazan, Russia}

\begin{abstract}
We report results of molecular dynamics simulations of liquid water
at the temperature $T=277$~K for a range of high pressure. One aim
of the study was to test the model Amoeba potential for description
of equilibrium structural properties and dynamical processes in
liquid water.  The comparison our numerical results with the Amoeba
and TIP5P potentials, our results of \emph{ab initio} molecular
dynamics simulations and the experimental data reveals that the
Amoeba potential reproduces correctly structural properties of the
liquid water. Other aim of our work  was related with investigation
of the pressure induced structural transformations and their
influence on the microscopic collective dynamics. We have found that
the structural anomaly at the pressure $p_c\approx 2000$~Atm is
related with the changes of the local, short-range order in liquid
water within first two coordination shells. This anomaly specifies
mainly by deformation of the hydrogen-bond network. We also discuss
in detail the anomalous behavior of sound propagation in liquid
water at high pressures and compare numerical results with the
experimental data.
\end{abstract}


\date{\today}
\end{frontmatter}

\section{Introduction}
Water is one of the most widespread liquids in wildlife and has a
great number chemical and technological applications
\cite{Franks_1972}. Although a particular water molecule has a
simple chemical structure, a water system is considered as a complex
fluid because of its anomalous behavior in thermodynamical,
structural and transport properties \cite{Stillinger_1980}. It dilates with
solidification, and the density has a maximum at the pressure
$1.0$~Atm and the temperature $277$~K \cite{Eizenberg_2005}.
Additionally, there is a minimum in the isothermal compressibility
at $319$~K and a clear minimum in the isobaric heat capacity at
$308$~K. These anomalies are linked with the microscopic structure
of liquid water \cite{Mokshin_2005}, which can be regarded as a \textit{transient gel}
-- a highly associated liquid with strongly directional hydrogen
bonds \cite{Geiger_1979,Stanley_1980}.

An active study of the different characteristics and properties of
water system is provided by the development of the numerical methods
of the computer simulations. So, the first study of the water
``structure'' was performed by Barker and Watts within Monte-Carlo
method \cite{Barker_1969}. They have computed the energy, heat
capacity and radial distribution function for water system at the
temperature $298$~K and have compared the obtained results with the
experimental data. The considered system was very small, $64$
molecules interacted via the Rowlinson
potential~\cite{Rowlinson_1951}. In the next study performed by
Rahman and Stillinger \cite{Rahman_1971} the water system was
simulated through $216$ rigid molecules interacted by means of the
pair-additive effective potential. The authors have investigated in
detail the structural, transport and dynamic properties of liquid
water at the temperature $307.3$~K.  The molecular dynamics
simulations in the low temperature thermodynamic phase range were
performed by Ref.~\cite{Poole_1992}. Here, the ``\emph{second
critical point}'' at the temperature $T=223$~K and pressure
$p=1000$~Atm was found. Below this second critical point, the liquid
phase separates into two distinct phases -- \textit{a low-density liquid} and
\textit{a high-density liquid}. It is necessary to note that to study the
water system a great number of the model potentials for intramolecular and
intermolecular interactions were suggested (see, for example,
review~\cite{Guillot_2002}). Here, the first model of the liquid
water was proposed in 1933 by Bernal and Fowler \cite{Bernal_1933}:
an ice-like disordered tetrahedral structure arising from the
electrostatic interactions between close neighbors.

The special role in the understanding of the microdynamic features
of the water have the methods of \textit{ab-initio} simulations based on the
density functional theory. So, one of the first works, where the
Car-Parrinello method was applied to study dynamic properties in
water, was done by Laasonen \emph{et al.}~\cite{Laasonen_1992}.
They also found that the local density approximation gives realistic
results for the intramolecular properties without gradient
corrections, but it fails to give a correct description of the
intermolecular interactions and the binding properties. Later, the
numerous \textit{ab-initio} simulations for water system have been done
\cite{Bernasconi_1998,Silvestrelli_1999,Schwegler_2000,Boero_2001,Kuo_2004}.

Recently, a large amount of the works were intended for the
investigation of water structural, transport, mechanical properties
at the different externally applied conditions
\cite{Lu_2008,Katayama_2010,Khan_2010,Han_2010,Gallo_2010}. For example,
the processes of the dissociation of the water molecules at the
pressures $14.5$~GPa and $26.8$~GPa  were investigated in
Ref.~\cite{Schwegler_2001} by means of \emph{ab-initio} molecular
dynamics simulations. Molecular dynamics simulations of water at the
negative pressures were carried out with two model potentials of
interparticle interaction: the extended simple point charge (SPC/E)
and the Mahoney-Jorgensen transferable intermolecular potential with
five points (TIP5P). The equilibrium phase diagram was constructed
in Ref.~\cite{Sanz_2004} for a wide range of temperature and
pressure by means of Monte-Carlo simulations within two model
potentials, TIP4P and SPC/E. Then, the correlation between
structural and dynamical anomalies in supercooled water was studied
\cite{Netz_2002} on the basis of the SPC/E potential.

Some of works  were pointed mainly towards the study of structural
and dynamical properties related with the anomalous features of
liquid
water \cite{Soper_2000a,Errington_2001,Saitta_2003,Li_2005,Esposito_2006,Yan2007,Krekelberg_2007,Krekelberg_2008,Debenedetti_2003}.
The concepts of low-density water (LDW) and high-density water (HDW)
were suggested in Ref.~\cite{Soper_2000a} on the basis of the
experimental data of neutron diffraction, which detected clear the
structural transformations in liquid water \cite{Errington_2001}.
Further, the structural transition from LDW to HDW as well as the
anomalies related with these structural transformations were
extensively
considered~\cite{Saitta_2003,Li_2005,Esposito_2006,Yan2007,Krekelberg_2007,Krekelberg_2008}.
An comprehensive overview of the properties of supercooled and
glassy water, the transition between low-density amorphous (LDA) ice
and high-density amorphous (HDA) ice were presented in
Ref.~\cite{Debenedetti_2003}. Nevertheless, in spite of the numerous
studies, the question about the impact of structural transition from
LDW to HDW on the dynamical properties and anomalies of water is
still open~\cite{Eizenberg_2005}.

In this work we perform the numerical study of the structural
transformations in water, which are appeared at the fixed
temperature $T=277$~K with the change of the externally applied
pressure $p=1.0 \div 10\;000$ Atm. In addition, the influence of the
structural rearrangements on the dynamical properties in liquid
water is also considered. Other aim of the work is related with the
test of the Amoeba potential~\cite{Ponder_2003,Ren_2003,Ren_2004} to
describe the structural and dynamical features of water.

The paper is organized in the following way. In the next section, we
describe the details of the molecular dynamics simulations of liquid
water on the basis of the Amoeba model potential. A detailed
analysis of the pressure dependence of the static properties and the
structure anomalies are then reported. Then, the features in the
dynamic and dispersion of sound velocity are reported and compared
with the experimental data on inelastic X-rays scattering.
The role of the hydrogen-bond network in the structural transformation
is discussed. And, finally, we conclude with a short summary.

\section{Computational Details \label{comp}}

Equilibrium classical molecular dynamics (MD) simulations of water
at the constant temperature $T=277$~K and the different pressures
were carried out with the TINKER molecular modeling package
\cite{Ponder_2003}. We use the Amoeba (Atomic Multipole Optimized
Energetics for Biomolecular Applications) water model, which was
recently suggested by Ren and Ponder \cite{Ren_2003,Ren_2004}. Our
computations were performed for $4000$ water molecules interacted
within a cubic box, where the periodic boundary conditions were
imposed in all directions. The Ewald summation was used to handle
the electrostatic interactions as well as an atom-based switching
window of the size 12~\textrm{\AA} was applied to cut off the van
der Waals interactions. The equations of motion were integrated via
a modified Beeman algorithm \cite{Allen_Tildesley} with the time
step $\Delta \tau = 1.0$~fs. The MD calculations were performed in
the isothermal-isobaric ensemble at the temperature $T=277$~K and the
pressure $p=1.0$, $1000$, $2000$, $3750$ $5000$, $7797$ and $10000$~
Atm, where we used the Berendsen thermostat and
barostat~\cite{Berendsen_1984} to enforce the constant temperature
and pressure.

\section{Results}

\subsection{Partial Radial Distribution Functions}

The radial distribution function (RDF) provides directly the
information about the average packing the molecules in water and can
be extracted from neutron \cite{Soper_1986} and X-ray scattering
data \cite{Hura_2000}. On the other hand, this term in an extended
form can be calculated on the basis of MD simulation data
\cite{Hansen/McDonald}, namely
\begin{equation}
g_{\alpha,\beta}(r)=\frac{L^3}{N_{\alpha}N_{\beta}}\Biggl\langle\sum_{j=1}^{N_{\alpha}}
\frac{n_{j\beta}(r)}{4\pi r^2 \Delta r}\Biggr\rangle, \label{Eq_gr}
\end{equation}
where $g_{\alpha,\beta}(r)$ is the probability to find an atom in
the range $[r,r+\Delta r]$ and $L$ is the length of the box edge.
The indexes $\alpha$ and $\beta$ define the type of the atom,
$\alpha,\beta \in \{O,H\}$, whereas $N_{\alpha}$ and $N_{\beta}$ are
the particle numbers of the type $\alpha$ and $\beta$, respectively.
The quantity $n_{j\beta}(r)$ specifies the number of $\beta$-atoms
located  at a distance $r$ from a $j$th atom within a spherical
layer of the thickness $\Delta r$.

\begin{figure}
\begin{center}
\includegraphics[width=1.15\textwidth]{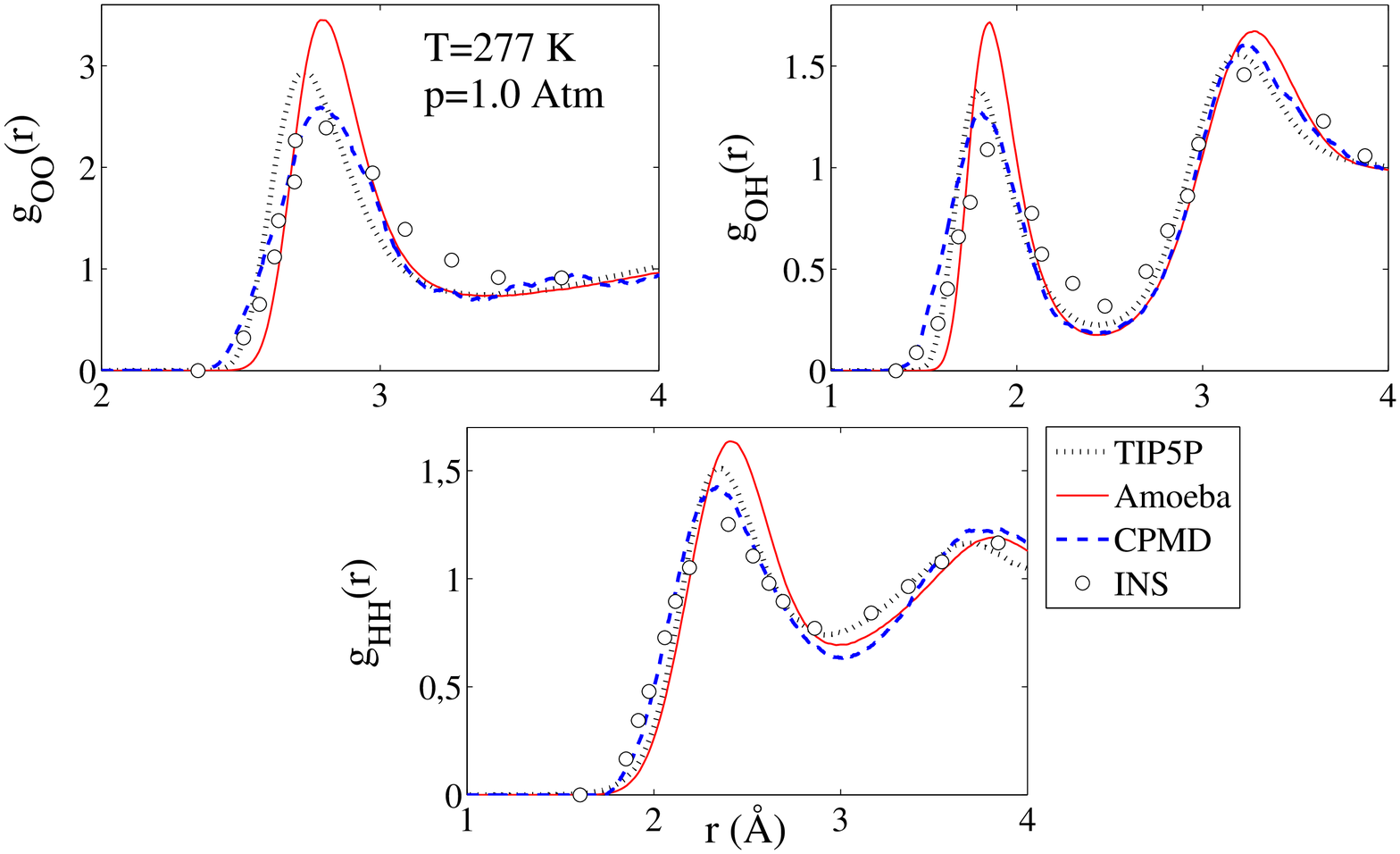}
\end{center}
\caption[kurzform]{\label{RDF_CMD_CPMD_INS} (color online) Partial
radial distribution functions of atoms $g_{\alpha,\beta}(r)$
($\alpha,\beta \in \{O,H\}$) at the temperature $T=277$~K and the
pressure $p=1.0$~Atm. Solid line represents the results of classical
molecular dynamics simulation with the Amoeba potential;
dotted line show the results of molecular dynamics simulations with
the TIP5P potential; dashed line represents the
results of Car-Parrinello molecular dynamics simulation for the
system with $64$ molecules \cite{CPMD}; circles are
the experimental data on neutron diffraction \cite{Soper_2000}.}
\end{figure}

In Fig.~\ref{RDF_CMD_CPMD_INS}, the partial RDF's of atoms
$g_{\alpha,\beta}(r)$ ($\alpha,\beta \in \{O,H\}$) obtained from our
simulations with the Amoeba potential at the temperature $T=277$~K
and the pressure $p=1.0$~Atm are compared with the molecular
dynamics simulations with the TIP5P potential, results of
\emph{ab-initio} Car-Parrinello molecular dynamics \cite{CPMD} and experimental
data on neutron diffraction~\cite{Soper_2000}. Although molecular dynamics
simulations data with the Amoeba potential overestimate
insignificantly the high of the first peak in partial RDF's in
comparison with experimental data and results of \emph{ab-initio}
molecular dynamics simulations, they reproduce correctly the
structure of liquid water at the temperature $T=277$~K and the
pressure $p=1.0$~Atm. Moreover, the comparison reveals that the
model TIP5P has a better agreement with experimental data and
results of \emph{ab-initio} molecular dynamics simulations.

\begin{figure}
\begin{center}
\includegraphics[width=1.2\textwidth]{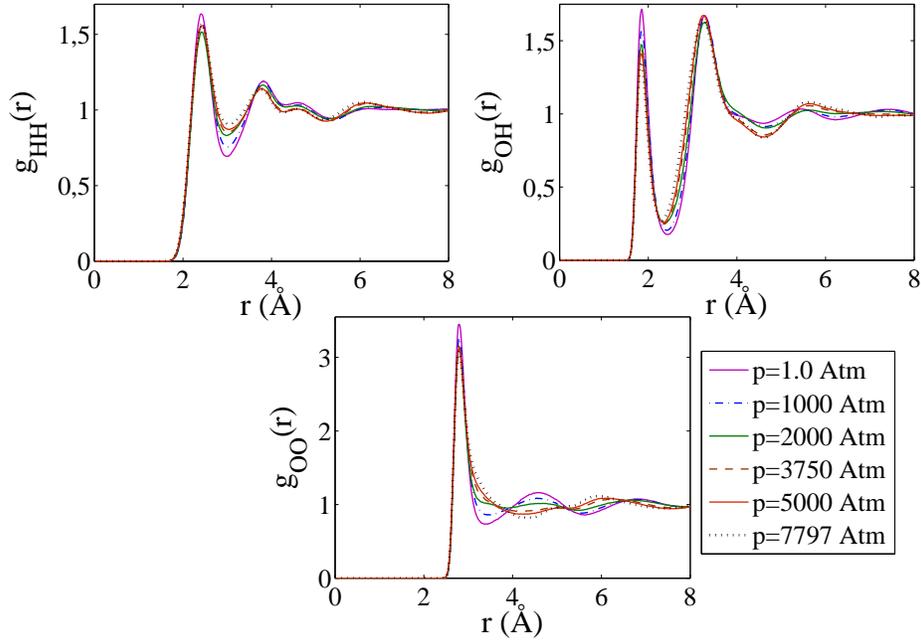}
\end{center}
\caption[kurzform]{\label{RDF} (color online) Partial radial
distribution functions of atoms $g_{\alpha,\beta}(r)$ ($\alpha,\beta
\in \{O,H\}$) at the temperature $T=277$~K and the different
pressures.}
\end{figure}

The pressure dependence of the partial RDF for water
is presented in Fig.~\ref{RDF}. As one can see
from the figure, the partial contributions of RDF for the pairs
$\textrm{H-H}$ and $\textrm{O-H}$ remain unchanged practically at
the increase of the pressure. The remarkable changes are observed in
the pressure dependence of RDF for O-O pairs and indicate clear on the
transition pressure $p_{tr} \sim 2000$~Atm. It is interesting that
the atomic rearrangements corresponding to the structural
transformation are restricted to the second coordination sphere,
where the shift of the second peak in $g_{\mathrm{OO}}(r)$ to the
range of the high $r$ occurs.

\subsection{Partial Static Structure Factors}

\begin{figure}
\begin{center}
\includegraphics[width=1.2\textwidth]{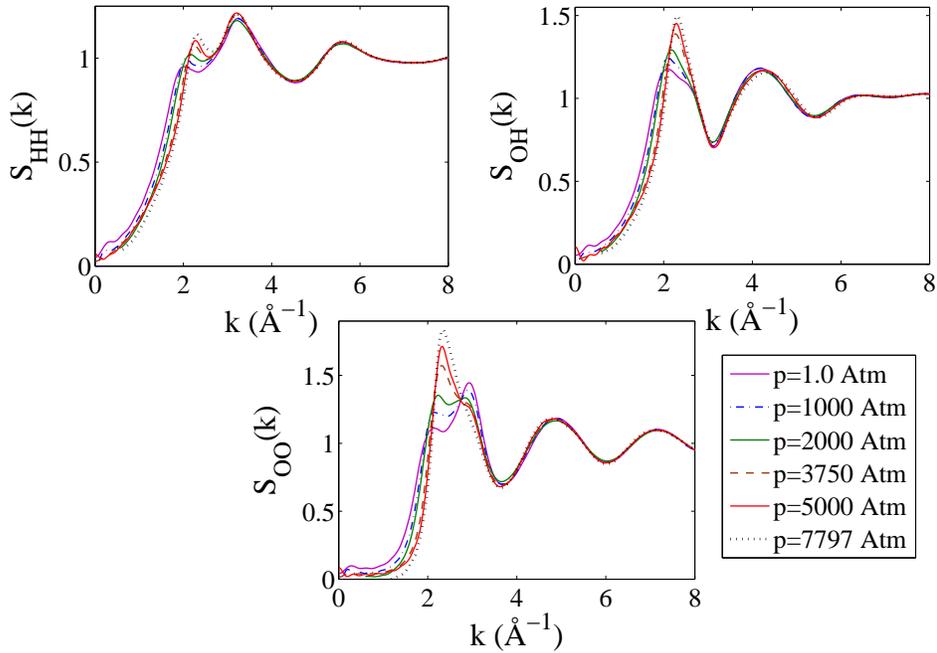}
\end{center}
\caption[kurzform]{\label{Sk} (color online) Partial static
structure factor $S_{\alpha,\beta}(k)$ ($\alpha,\beta \in \{O,H\}$)
at the temperature $T=277$~K and the different pressures.}
\end{figure}

Other term, which is sensitive to the structural transformations, is
the static structure factor, $S(k)$. This term is measured
experimentally and is defined as \cite{Hansen/McDonald}
\begin{equation}
S(k)=\frac{1}{N}\left\langle
\sum_{j=1}^{N}e^{-i\textbf{kr}_{j}}\sum_{l=1}^{N}e^{i\textbf{kr}_{l}}\right\rangle.
\label{Eq_Sk}
\end{equation}
Here, $\textbf{k}$ is the wave-vector and  $k = |\textbf{k}|$;
$\textbf{r}_j(t)$ defines the radius-vector of an $j$th atom. The
partial static structure factor $S_{\alpha,\beta}(k)$, as an
extension of $S(k)$, is directly related with the Fourier transform
of the radial distribution function \cite{Allen_Tildesley}:
\begin{equation}
S_{\alpha,\beta}(k)=1+4\pi
n\int_{0}^{\infty}r^{2}\Bigg[g_{\alpha,\beta}(r)-1\Bigg]\frac{\sin(kr)}{kr}dr,
\label{Eq_Sk_par}
\end{equation}
where $n=N/V$ is the numerical density.

The partial static structure factor $S_{\alpha,\beta}(k)$ at various
pressures for different atom pairs, $\textrm{H-H}$, $\textrm{O-H}$ and
$\textrm{O-O}$, is presented in Fig.~\ref{Sk}. As can be seen, the
features in the first peak of $S_{\alpha,\beta}(k)$ become more
pronounced with the increase of pressure that indicates on the rise
of the short-range order. Further, the split of the first peak in
$S_{\mathrm{OO}}(k)$ disappears with the increase of pressure that
is related with the local molecular ordering in the vicinity of the
first coordination shell.

\subsection{Wendt-Abraham Parameter, Pair-Correlation Entropy and Translational Order Parameter}
An additional information about the local structural features of the
system can be extracted with the help of the extended Wendt-Abraham
parameter
\begin{equation}
r_{\alpha,\beta}^{WA}=\frac{g_{\alpha,\beta}^{min}(r)}{g_{\alpha,\beta}^{max}(r)},
\label{Eq_WA}
\end{equation}
where $g_{\alpha,\beta}^{max}(r)$ and $g_{\alpha,\beta}^{min}(r)$
represent the main maximum and the first minimum in a partial radial
distribution function, respectively~\cite{Wendt_Abraham}.
\begin{figure}
\begin{center}
\includegraphics[width=1.95\textwidth]{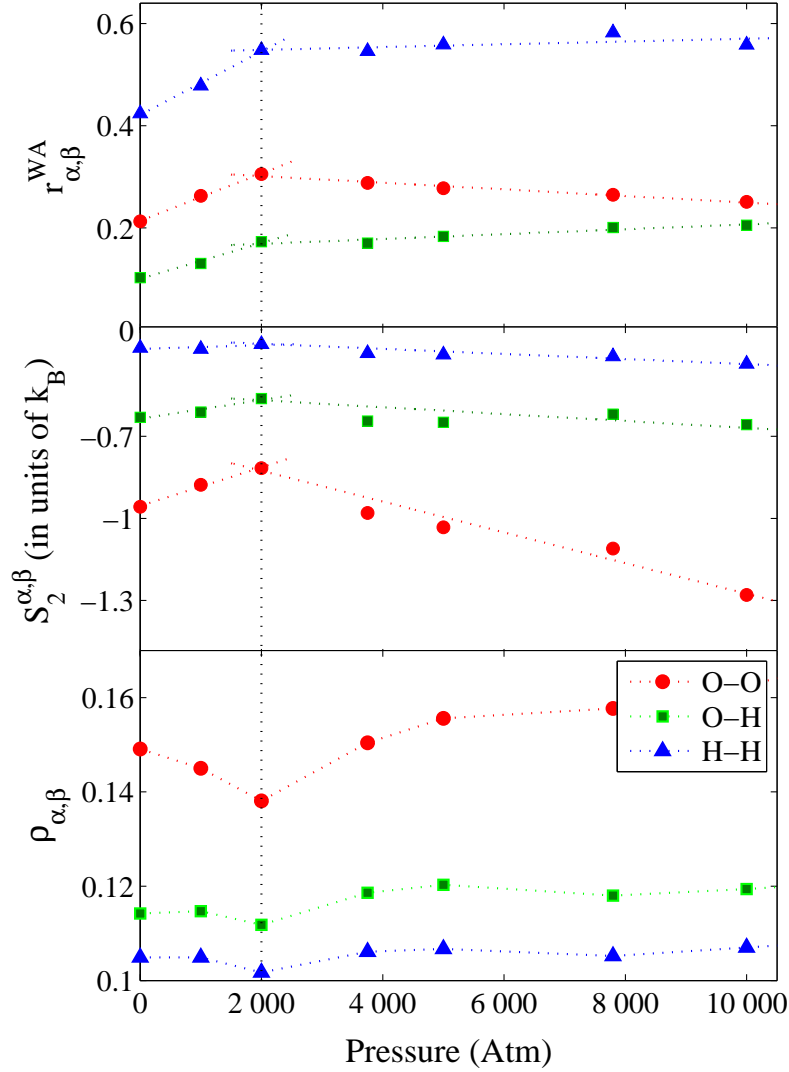}
\end{center}
\caption[kurzform]{\label{Fig_WA} (color online) Pressure dependence
of the partial Wendt-Abraham parameter $r_{\alpha,\beta}^{WA}$, the
partial pair-correlation entropy $S_2^{\alpha,\beta}$ and the
partial translational order parameter $\rho_{{\alpha,\beta}}$, where
$\alpha,\beta \in \{O,H\}$.}
\end{figure}
The thermodynamic excess entropy of a fluid is defined as the
difference in entropy between the fluid and the corresponding ideal
gas under identical temperature and density conditions. The total
entropy of a classical fluid can be written as
\begin{equation}
S=S_{id}+\sum_{n=2}^NS_n,
\label{Eq_S}
\end{equation}
where $S_{id}$ is the entropy of the ideal gas reference state,
$S_n$ is the entropy contribution due to $n$-particle spatial
correlations. Then, the excess entropy is defined as
\begin{equation}
S_{ex}=S-S_{id}.
\label{Eq_S_ex}
\end{equation}
The main contribution of $S_{ex}$ for classical fluids is a
pair-correlation entropy $S_2$ (for example, in case of a monoatomic
liquids this two-particle contribution to the excess entropy is
$85\div95 \%$ over a fairly wide range of densities), which can
be given in the extended form as
\begin{equation}
S_2^{\alpha,\beta}=-2\pi n \int_0^{\infty}\bigg \{
g_{\alpha,\beta}(r)\ln(g_{\alpha,\beta}(r))-[ g_{\alpha,\beta}(r)-1
]\bigg\} r^2 dr,
\label{Eq_S2}
\end{equation}
where the term $S_2$ is measured in units of $k_B$.

The translational order parameter $\rho$ is a simple and convenient
measure to test the radial ordering in the studied system
\cite{Tanaka_1998}. The separation of $\rho$ into partial
contributions can be written as the following
\begin{equation}
\rho_{\alpha,\beta} =\frac{1}{r_m}\int_{0}^{r_m}|g_{\alpha,\beta}(r)-1|dr,
\label{Eq_top}
\end{equation}
where $r_m$ is the distance at which the structural features of the
system cease to be considered. The value of $r_m$ corresponds to a
half of the edge of the simulation box. It is clear that an
increase of $\rho_{\alpha,\beta}$ indicates directly on the growth
of structural radial ordering, whereas the increase of
$r_{\alpha,\beta}^{WA}$ and $S_2^{\alpha,\beta}$ manifests the
processes of structural disordering in the system.

Figure \ref{Fig_WA} illustrates the pressure dependence of the
partial Wendt-Abraham parameter $r_{\alpha,\beta}^{WA}$, the partial
pair-correlation entropy $S_2^{\alpha,\beta}$ and the partial
translational order parameter $\rho_{{\alpha,\beta}}$ at the fixed
temperature $T = 277$~K. The decrease of $r^{WA}$ indicates on the
increase of the two-particle chain clusters in the system, which
appear, e.g., due to the system densification and precedes the
transition of the system into a solid (crystalline or amorphous)
phase. As can be seen from the figure, all parameters discover the
change in the pressure dependence at $p_c \approx 2000$~Atm. It is
remarkable that the extended Wendt-Abraham parameter
$r_{\alpha,\beta}^{WA}$ is more sensitive to detect these structural
transformations than other parameters, $S_2^{\alpha,\beta}$ and
$\rho_{\alpha,\beta}$. This can be considered directly as an
evidence that the structural transformations detected at $p_c
\approx 2000$~Atm are related with the changes of the local,
short-range order, since the parameter $r_{\alpha,\beta}^{WA}$
probes the structural properties inside of the first coordination
shell whereas the quantities $S_2^{\alpha,\beta}$ and
$\rho_{\alpha,\beta}$ characterize features at more extended spatial
scales.

\subsection{Tetrahedral Order Parameter}
The tetrahedral order parameter $Q_k$ can be used to quantify the
tetrahedrality within the first shell~\cite{Errington_2001}. This
parameter is defined as
\begin{equation}
Q_k=1-\frac{3}{8}\sum_{i=1}^3\sum_{j=i+1}^4 \bigg[\cos \theta_{ikj}+\frac{1}{3} \bigg]^2.
\label{Eq_Q_k}
\end{equation}
The term $\theta_{ikj}$ is the angle formed between the central
molecule $k$ and its neighbors $i$ and $j$. The average value
\begin{equation}
\langle Q \rangle=\frac{1}{N}\sum_{k=1}^N Q_k
\label{Eq_Q}
\end{equation}
quantifies directly the orientational order within the first shell.
For the perfect tetrahedral order, one has $\langle Q \rangle=1$; if
the bonds are arranged in a manner with the absence of tetrahedral
geometry, one has $\langle Q \rangle=0$. So, with the deviation from
the perfect tetrahedron the values of $\langle Q \rangle$
demonstrate decease.

\begin{figure}
\begin{center}
\includegraphics[width=0.8\textwidth]{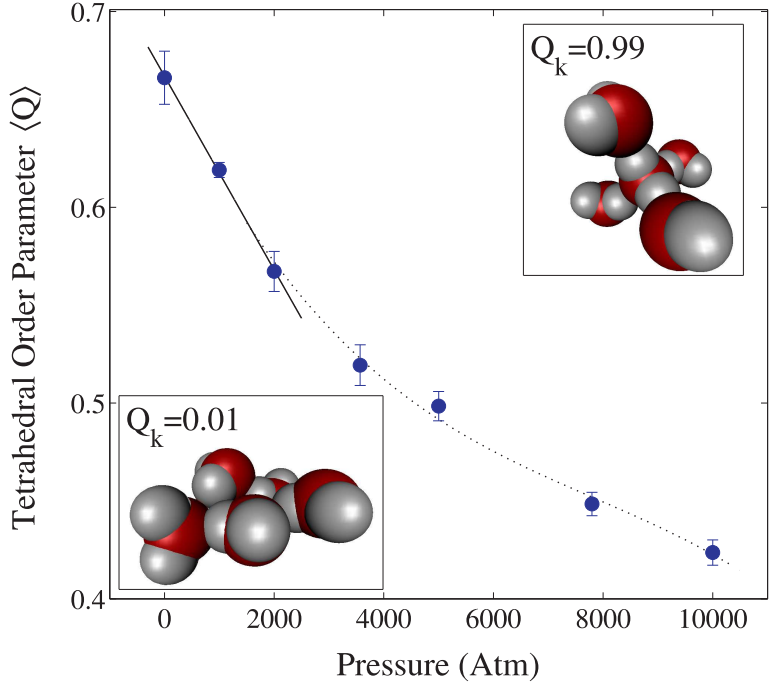}
\end{center}
\caption[kurzform]{\label{TetOrdPar} (color online) Main:
Tetrahedral order parameter \emph{versus} pressure at the
temperature $T=277$~K. Inset: Instantaneous configurations of water
molecules at the two values of the tetrahedral order parameter:
$Q_k=0.01$ (the four bonds are practically aligned) and  $Q_k=0.99$
(molecular system is close to be a perfect tetrahedron).}
\end{figure}

The pressure dependence of the tetrahedral order parameter $\langle
Q \rangle$ for liquid water at the temperature $T=277$~K is
presented in Fig.~\ref{TetOrdPar}. As can be seen, with the increase
of the applied pressure the  tetrahedral order parameter
demonstrates a nonlinear decrease. Such a behavior of the parameter
indicates on the decrease of the amount of tetrahedral structures in
liquid water with the isothermal increase of the pressure $p$, that
is in agreement with the results of Ref.~\cite{Soper_2000a}. It is
necessary to note that although the behavior of the tetrahedral
order parameter $\langle Q \rangle$ changes with the pressure $p$,
this parameter $\langle Q \rangle$ demonstrates no a sharp
transition from low-density phase to high-density one.

\subsection{Dynamic Structure Factor and Dispersion of Sound Velocity}
According to a definition (see Ref.~\cite{Hansen/McDonald}), the
dynamic structure factor $S(k,\omega)$ is related with the Fourier
transform of the intermediate scattering function $F(k,t)$
\begin{equation}
S(k,\omega)=\frac{1}{\pi}\int_0^{\infty}F(k,t)\cos(\omega t)dt.
\label{Eq_DSF}
\end{equation}
Here
\begin{eqnarray}
F(k,t)=\frac{1}{N}\Bigg\langle \sum_l e^{i\textbf{k}\cdot \textbf{r}_l(0)}
\sum_j e^{-i\textbf{k}\cdot \textbf{r}_j(t)}\Bigg\rangle=
\big\langle \delta \rho_k(0)\delta \rho^*_k(t) \big\rangle,
\label{Eq_ISF}
\end{eqnarray}
where $\delta \rho_k(t)$ is the $k$-component of the fluctuation of
the microscopic number density $\rho_k(t)$. Nevertheless, because of
the long-time tails of $F(k,t)$ in water (see
Ref.~\cite{Harder_2005}), the finding of the dynamic structure
factor spectra $S(k,\omega)$ by means of Eqs.~(\ref{Eq_DSF}) and
(\ref{Eq_ISF}) yields the significant inaccuracies. To avoid this
problem we use the Wiener-Khinchin (statistical-autocorrelation)
theorem \cite{Korn/Korn}, which allows one to rewrite Eq.
(\ref{Eq_DSF}) in the next identical form
\begin{equation}
S(k,\omega)=\frac{1}{t_M}\Bigg|\int_{0}^{t_M}\delta \rho_k(t)e^{i\omega t}dt\Bigg|^2,
\label{DinStrFact}
\end{equation}
where $t_M$ is the observation time for the variable $\delta
\rho_k(t)$. Eq. (\ref{DinStrFact}) gives a possibility to find
$S(k,\omega)$ from the dynamical variable $\delta \rho_k(t)$,
directly. Note that an advantage of Eq. (\ref{DinStrFact}) over Eq.
(\ref{Eq_DSF}) is related with the saving of computational time,
that is very important at numerical estimations. Then, the dynamic
structure factor $S(k,\omega)$  can be evaluated at these
$k$-values, which are allowed by the system size: $k=2\pi/L
(n,m,l)$, where $n,\;m,\;l$ are integers. The size of our system
allows one to take the smallest value of $k \approx 1.38$ nm$^{-1}$.
\begin{figure}
\includegraphics[width=1.25\textwidth]{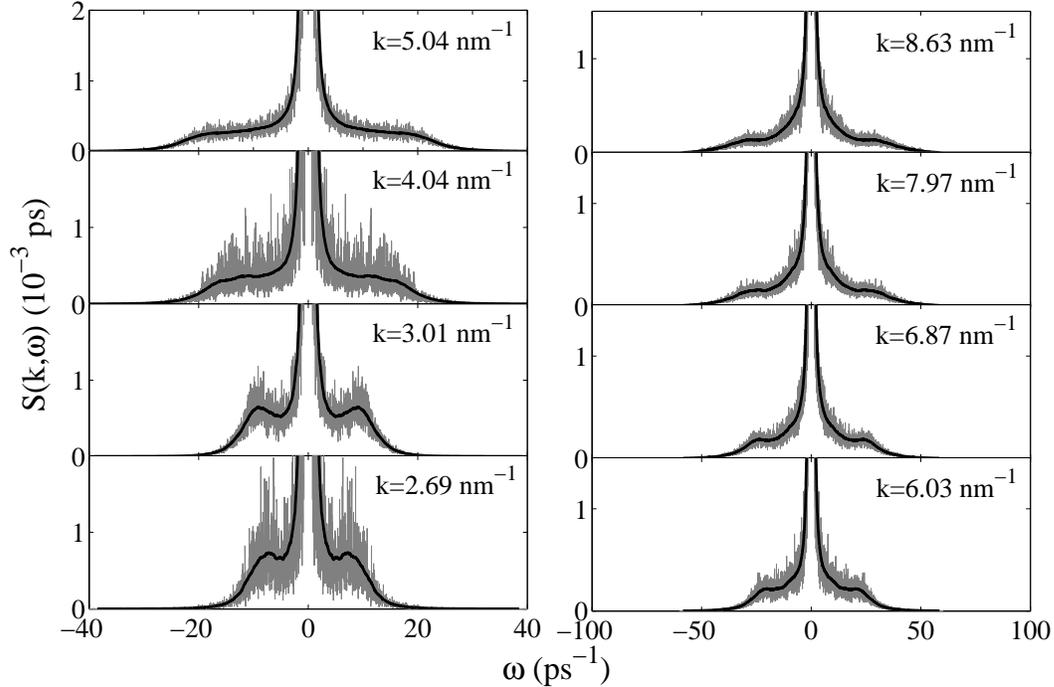}
\caption[kurzform]{\label{DSF} (color online) Dynamic structure
factor of liquid water at the temperature $T=277$~K and the pressure
$p=3750$~Atm for different values of the wave number $k$: the noisy
(gray) lines represent the results obtained from molecular dynamics
simulation data and Eq.~(\ref{DinStrFact}), thick (black) lines are
results of the smoothing procedure.}
\end{figure}
To delete the noises from dynamic structure factor spectra
$S(k,\omega)$ we apply the smoothing procedure based on the
computation of the arithmetic mean for the local window with the
length $1.0$~ps$^{-1}$.

The frequency spectra of the dynamic structure factor $S(k,\omega)$
of water for different values of the wave number
$k=2.69\div8.63$~nm$^{-1}$ at the temperature $T=277$~K and the
pressure $p=3750$~Atm (that is higher than $p_c$) are presented in
Fig.~\ref{DSF}. Here, the high frequency collective excitations
responsible for sound propagation in the system are clearly detected
for all the considered values of $k$. To obtain the precise values
of the sound velocity $\omega_c(k)$ which is related with the
high-frequency peak-position in $S(k,\omega)$, we consider the
longitudinal current spectra $C_L(k,\omega)$ \cite{Ruocco_1999}.
This term can be found from the dynamic structure factor
$S(k,\omega)$ by the way:
\begin{equation}
C_L(k,\omega)=\frac{\omega^2}{k^2}S(k,\omega).
\label{LongCurSpec}
\end{equation}

\begin{figure}
\begin{center}
\includegraphics[width=1.1\textwidth]{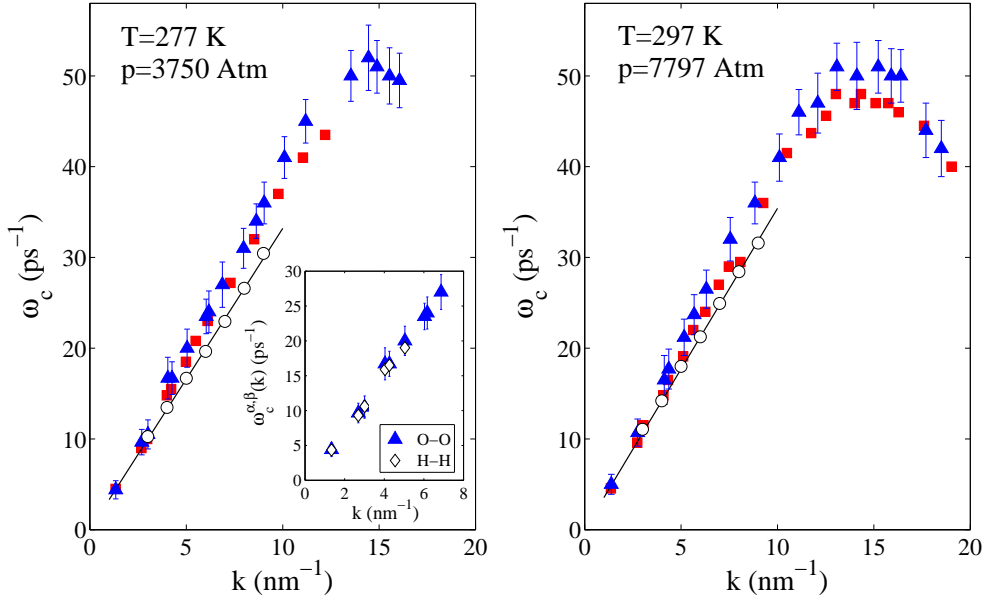}
\end{center}
\caption[kurzform]{\label{Dispers} (color online) Main: Dispersion
of sound velocity $\omega_c(k)$: circles represent the experimental
data \cite{Krisch_2002}; triangles are results of molecular dynamics
simulations data with Amoeba potential; squares represent results of
computer simulations with TIP5P potential; solid
line indicates the best linear fit to experimental data, and the
slope corresponds to the sound velocity $\vartheta=3320$~$m/s$
($\vartheta=3540$~$m/s$) at the temperature $T=277$~K ($T=297$~K).
Inset: Partial dispersion of sound velocity
$\omega^{\alpha,\beta}_c(k)$.}
\end{figure}

\begin{figure}
\begin{center}
\includegraphics[width=1.2\textwidth]{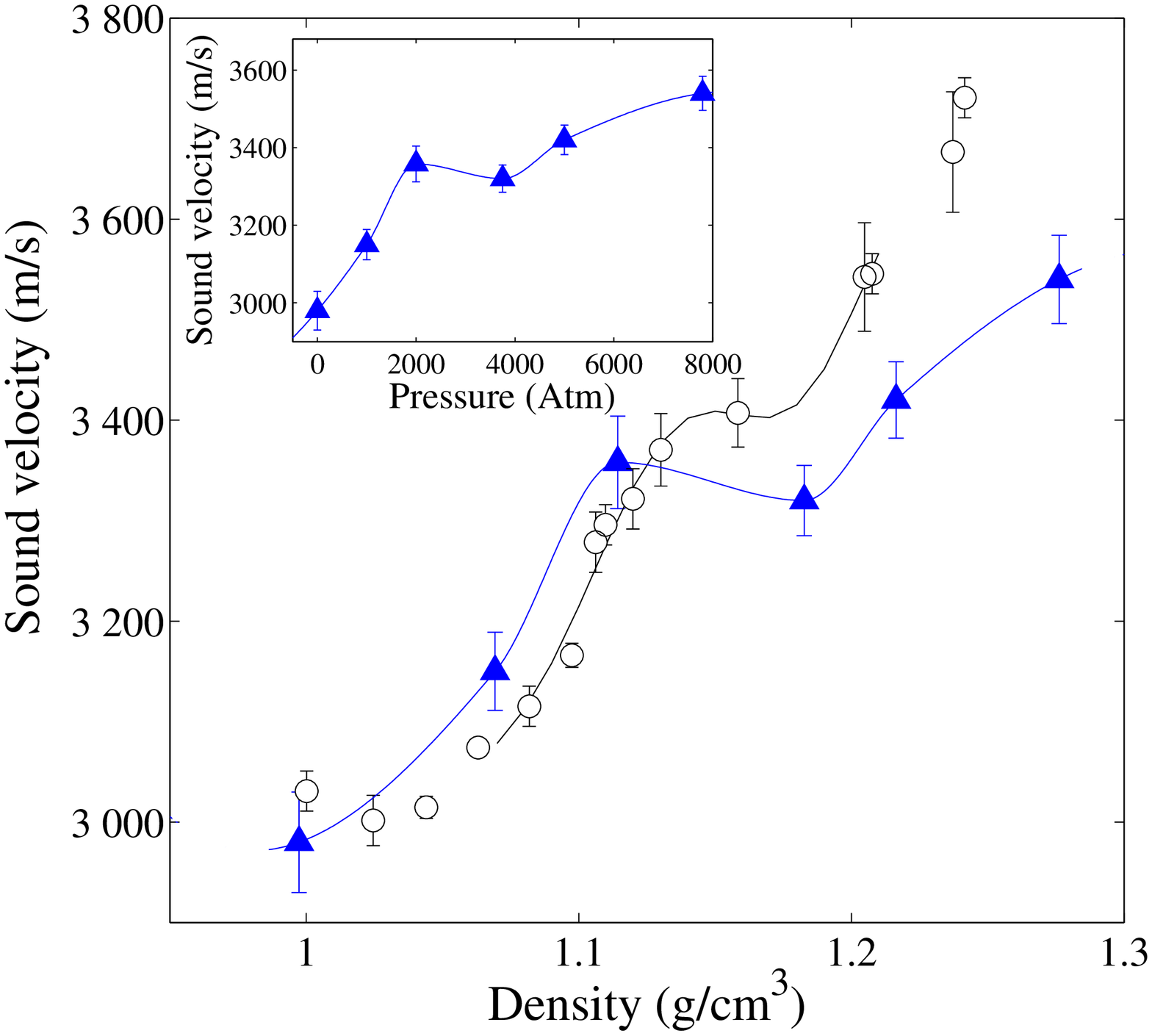}
\end{center}
\caption[kurzform]{\label{Sound} (color online) Main: Density
dependence of the sound velocity $\vartheta$: the circles represent
the experimental IXS data \cite{Krisch_2002}, the triangles are
results of our molecular dynamics simulations with Amoeba water
model. Solid line is the spline-interpolation. Inset: Pressure
dependence of the sound velocity obtained from simulations.}
\end{figure}
On Fig.~\ref{Dispers}, the dispersion of sound velocity
$\omega_c(k)$ in water at the temperatures  $T=277$~K and $T=297$~K
and the pressures $p=3750$~Atm and $p=7797$~Atm are presented.
Results of molecular dynamics simulations with the Amoeba model
potential reproduce correctly the experimental inelastic X-ray
scattering data \cite{Krisch_2002}. Moreover, with the increase of
$k$ the simulations results overestimate slightly the experimental
data for sound velocity of Ref.~\cite{Krisch_2002}. The comparison
of the results with the two potentials reveals that data obtained on
the basis TIP5P are in the range of inaccuracies of the dispersion
curve obtained on the basis Amoeba potential and have a better
agreement with experimental data. Nevertheless, simulation data with
TIP5P demonstrate also overestimated values in comparison with the
experimental outcomes of inelastic X-ray scattering.

The partial dispersion of sound velocity
$\omega^{\alpha,\beta}_c(k)$ for $O$ and $H$ atoms obtained with the
Amoeba potential is given in inset of Fig.~\ref{Dispers}. As can be
seen, both lines are identical. This is direct evidence that the
main influence on the features of sound propagation in water appear
due to oxygen, which is the heavy component of a water molecule.
Note that this is agreed with results of Ref.~~\cite{Ricci_1989}
(p.~7235).

To estimate quantitatively the pressure dependence on the properties
of the collective dynamics in water we show in Fig.~\ref{Sound} the
density and pressure dependence of the sound velocity $\vartheta$,
where results of our molecular dynamics simulations data are
compared with experimental data~\cite{Krisch_2002}. First, as can be
seen from figure, results of the molecular dynamics simulations
reproduce correctly the experimental data for sound
velocity~\cite{Krisch_2002}. An insignificant understatement of
$\vartheta$ appears from simulations in comparison with experimental
data at high densities. Nevertheless, both our numerical results and
experimental inelastic X-ray scattering data reveal clearly a
changes in dynamics at the density $\rho_c =1.114 \div 1.12$
g/cm$^3$.  Note that such density of the investigated water system
corresponds to the pressure $p_c\approx 2000$ Atm (see inset of
Fig.~\ref{Sound}), at which the structural transformations were
detected by the order characteristics.

Thus, the structural and dynamic anomalies observed in water at the
pressure $p_c \approx 2000$~Atm are provided by the local structural
molecular rearrangement inside of the first two coordination shells.
It is necessary to note that such a treatment is also suggested by
results of Ref.~\cite{Yan2007}, where anomalies in water were
studied on the basis of the five-site transferable interaction
potential (TIP5P). It was found in Ref.~\cite{Yan2007} that the
anomalous decrease of orientational order upon compression occurs in
the first and second coordination shells, but the anomalous decrease
of translational order upon compression occurs mainly in the second
shell.

\subsection{Hydrogen-Bond Network}
A detailed description of the hydrogen-bond network in liquid water is the key
to understanding its unusual properties.
The studies of the hydrogen-bonded network structure in water have mainly
relied on neutron and X-ray diffraction, infrared spectroscopies, and
computer simulations techniques \cite{Ludwig_2001}. Experimental data from noncrystalline
materials provide the radial distribution functions and static structure factors \cite{Soper_2000}
that do not provide angular correlations needed to assign uniquely the local
geometries in water \cite{Kusalik_1994}.
\begin{figure}
\begin{center}
\includegraphics[width=0.8\textwidth]{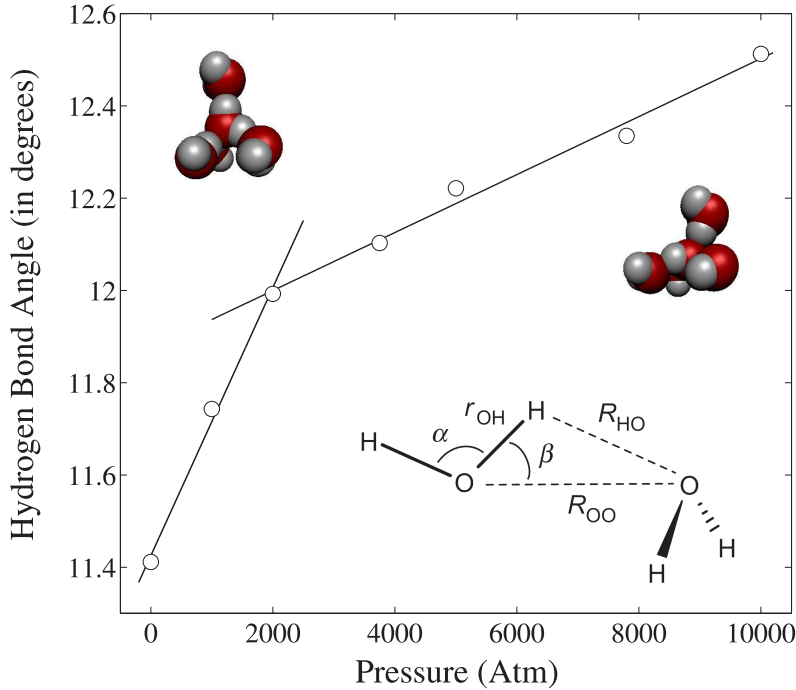}
\end{center}
\caption[kurzform]{\label{H_Bond_Angle} (color online) Pressure dependence of the
hydrogen bond angle.}
\end{figure}
In order to understand the effect of the structure on the dynamics we
carried out a detailed analysis of the pressure dependence of the
hydrogen bond angle \cite{Chaplin,Modig_2003}.
In the same lines with the work \cite{Luzar_1993}, we use a
\textit{geometric criterion}. According to this criterion two water
molecules are taken to be hydrogen-bonded if their inter-oxygen distance
is less than $3.5~ \textrm{\AA}$ and simultaneously the hydrogen-oxygen distance
is less than $2.45~ \textrm{\AA}$ and the oxygen-oxygen-hydrogen angle $\beta$ is
less than $30^{\circ}$. We note that the critical distances of $3.5~ \textrm{\AA}$
and $2.45~ \textrm{\AA}$ are essentially the positions of the first minimum in the
oxygen-oxygen and oxygen-hydrogen radial distribution functions, respectively.
All the terms of this algorithm are presented in the bottom inset of Fig.~\ref{H_Bond_Angle}.
The behavior of the hydrogen bond angle $\beta$ at the temperature $T=277$~K is
illustrated in Fig.~\ref{H_Bond_Angle}. This quantity is presented as an averaged
over all possible hydrogen-bond environments.
As can be seen from the figure, the distinct changes in the behavior of the angle
$\beta$ at critical pressure $p=2000$~Atm is observed.
In the inset of Fig.~\ref{H_Bond_Angle}, we show the configurations of
water molecules at the applied pressure $p=1.0$~Atm (on the left of Fig.~\ref{H_Bond_Angle}) and
$p=10000$~Atm (on the right of Fig.~\ref{H_Bond_Angle}) in the neighborhood of four
closest molecules, which are connected by hydrogen bonds.
Thereby, we confirm quantitatively the idea that the structural and dynamical
anomalies of water are related with the local structural rearrangement of molecules
within first two coordination shells \cite{Yan2007}, which specifies by
deformation of the hydrogen-bond network.

\section{Conclusions}

Molecular dynamics simulations of liquid water at the temperature
$T=277$~K and the range of pressure $p=1.0\div10\;000$ Atm are
performed on the basis of model Amoeba potential~\cite{Ponder_2003}.
It is found that this potential reproduces correctly equilibrium
structural and dynamical properties of liquid water, although TIP5P
potential yields a better agreement with experimental data.

The pressure dependence of the structural characteristics and order
parameters detects clearly the appearance of the structural anomaly
\cite{Yan2007} in the water system at the temperature $T=277$~K  and
the pressure $p\approx 2000$~Atm. The detailed analysis of the
extended Wendt-Abraham parameter, pair-correlation entropy and
translational order parameter indicates that the observed structural
transformations occur in the hydrogen/oxygen atoms in the range of
the first two coordination shells. Thus, these structural
transformations are correlated with the deformation of the
hydrogen-bond network \cite{Modig_2003}. Moreover, it is found that
with the increase of the applied pressure, a fraction of tetrahedral
structures in the system decreases. The tetrahedral order parameter
$\langle Q \rangle$ demonstrates no a sharp transition from
low-density state to high-density one.

We have found that the structural anomaly has an impact on the
dynamic properties of the liquid water. The computed spectra of the
dynamic structure factor $S(k,\omega)$ at high pressures reveal all
features (including the high-frequency acoustic mode) typical for
liquid systems \cite{Mokshin_2005,Yulm_2001,Scopigno_RMP}. The found
sound dispersions are in a good agreement with the experimental
inelastic X-ray scattering data~\cite{Krisch_2002}. It is found the
main features of sound propagation in liquid water are related with
the oxygen atom.

The pressure dependence of the sound velocity
indicates on the anomalous dynamics in water at the same value of
the pressure as for the structural anomaly, $p\approx 2000$ Atm (the
temperature $T=277$~K), that corresponds to the density $\rho_c
\approx 1.12$ g/cm$^3$.

\section{Acknowledgments}
We dedicate this paper to the memory of Professor Renat M.
Yulmetyev. This work was supported by the grants of Russian
Foundation for Basic Research (No. 08-02-00123-a and No.
09-02-91053-CNRS-a).

\end{document}